\documentclass{kluwer}    % Specifies the document style.

\usepackage{epsfig}
\newdisplay{guess}{Conjecture}

\begin{document}                                                          
\begin{article}
\begin{opening}         
\title{Merging Rate of Dark Matter Halos: \\ Evolution
and Dependence on Environment} 
\author{Stefan \surname{Gottl\"ober}}  
\runningauthor{Stefan Gottl\"ober et al.}
\runningtitle{Merging rate of halos}
\institute{Astrophysical Institute Potsdam, An der Sternwarte 16, D-14482
Potsdam, Germany}
\author{Anatoly \surname{Klypin}}  
\author{Andrey V.~\surname{Kravtsov}}  
\institute{Astronomy Department, NMSU, Dept.4500, Las Cruces, NM
88003-0001, USA}

\begin{abstract}
We discuss the impact of the cosmological environment on the evolution
of dark matter halos using a high-resolution simulation within a 
spatially flat $\Lambda$CDM cosmology. 
\end{abstract}

\end{opening}           

\section {Introduction}

It is generally believed that cold dark matter (DM) dominates the mass
in the Universe and significantly affects both the process of galaxy
formation and the large-scale distribution of galaxies. Here we
present results of a study of the formation and the evolution of the
DM component of galaxies, DM halos. The structure of the halos depends
on the environment (e.g., Avila-Reese et al. 1999), so that the
properties of galaxies are also expected to depend on the cosmological
environment.  For our analysis we use a low-density flat cosmological
model with cosmological constant $\Lambda$, which have been proved to
be very successful in describing most of the observational data at
both low and high redshifts: $\Omega_M=1-\Omega_{\Lambda}=0.3$,
$\sigma_8=1$, $H_0=70$ km/s/Mpc, $t_0 \approx 13.5$ Gyrs.

\section {Numerical Simulations}

In order to study the statistical properties of halos and to have 
a sufficient mass resolution we have chosen a simulation box of $60
h^{-1} {\rm Mpc}$ with $256^3$ cold dark matter particles (particle
mass of $1.1 \times 10^9 h^{-1} {\rm M_{\odot}}$).  Using the Adaptive
Refinement Tree (ART) code (Kravtsov, Klypin \& Khokhlov 1997) we
reached a force resolution of $\approx 2h^{-1}$ kpc in high density
regions.

Identification of halos in dense environments and reconstruction of
their evolution is a challenge. Any halo finding algorithm has to deal
with difficult ``decision-making'' situations, in particular when many
gravitationally bound halos are moving within a large dense object
(a galaxy cluster or a group). We use an algorithm described
in Klypin et al. (1999).  We characterize each halo by its mass and
the maximum
circular velocity $v_{circ}=\sqrt{GM/R}$. This quantity is more
meaningful observationally and can be numerically measured more easily
and more accurately than mass.  The halo samples are complete for halos
with $v_{circ} {_ >\atop{^\sim}} 100$ km/s (Gottl\"ober et
al. 1999). Here we consider only more massive halos with circular
velocities $v_{circ} > 120$ km/s. The limit on the circular velocity is
relaxed at high redshifts to allow identification of small progenitors
of present-day halos. At $z=0$ we have detected 4193 halos in our
simulation, which corresponds to a halo number density of $0.019 h^3
{\rm Mpc}^{-3}$.

For each halo in our $z=0$ sample we have constructed a complete
evolution tree over up to 23 time moments approximately regular in time
distributed between $z = 0$ and $z = 10$.  The procedure of progenitor
identification is based on the comparison of lists of particles
belonging to the halos at different moments both back and forward in
time (Gottl\"ober et al 1999). The epoch at which the halo has been
identified for the first time depends on our assumption of thresholds
of the circular velocity ($v_{circ}>50$km/s) and mass (minimum of 40
bound particles) for the progenitors of our halos at $z>0$.

\section {Environment of Halos}

In order to find the cosmological environment of each of the 4193 halos
we run a friend-of-friend analysis over the dark matter particles with
a linking length of 0.2 times the mean interparticle distance. By this
procedure we find clusters of dark matter particles with an overdensity
of $\approx 200$. The virial overdensity in the $\Lambda$CDM model under
consideration is $\approx 330$, which corresponds to a linking length of
0.17. Therefore, the objects which we find are slightly larger
than the objects at the virial overdensity. We have increased the
linking length because we found that halos, which are outside, but
close to a cluster, are affected by the cluster.

For each of the halos at $z=0$ we find the cluster of dark matter
particles to which the halo belongs. We call the halo a cluster galaxy
if the halo belongs to a particle cluster with a total mass larger than
$10^{14} h^{-1} {\rm M_{\odot}}$. We call it an isolated galaxy if only
one halo belongs to the object at overdensity 200. The rest of halos
are called group galaxies. By this definition pairs of galaxies are
also handled as groups. This increases slightly the number of galaxies
in ``groups''.

With the procedure described above we find at $z = 0$ that there are
401 cluster galaxies (9.6\%), 1247 galaxies in groups (29.7\%), and
2545 isolated galaxies (60.7\%). If we consider objects of virial
overdensity, the number of cluster and group galaxies decreases to
6.5\% and 25.6\% and 67.9\% of galaxies become ``isolated''. For
comparison see also Avila-Reese et al. (1999).

\section {Merging Rate of Halos}

The mass of a halo increases due to accretion and
merging. However, interacting halos may exchange and lose 
mass.  We calculate the relative mass growth per time
interval $(M_1 - M_2)/M_2/(t_2-t_1)$, where time is in units of
$10^9$ years. If it is larger than 0.35, we call this a major-merging
event. Note that according to this definition, we calculate the {\it
total} change of mass, not merging with a large halo.

\begin{figure} 
\centerline{\epsfig{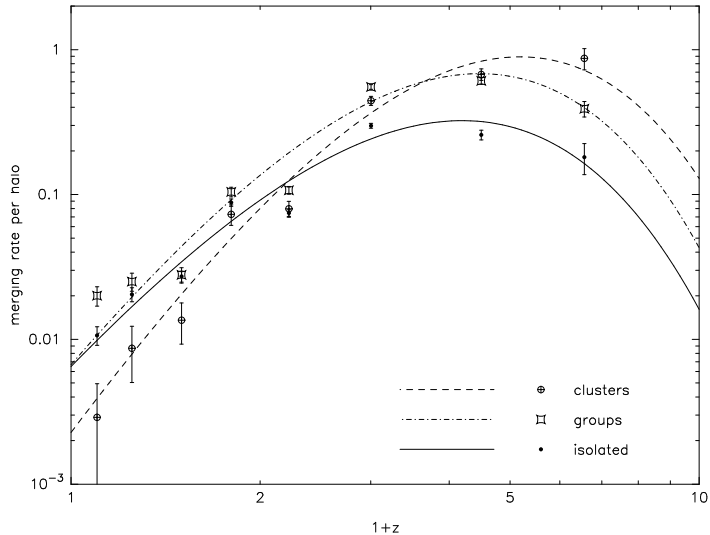}}
\caption[]{Merging rate per halo and Gyr for isolated halos, halos
in clusters, and halos in groups (different symbols). Curves are
analytical fits of the form $(1+z)^{\beta}\exp(-0.6(1+z))$. Halos in
clusters had much higher merging rates at high redshifts well before the
formation of clusters. At $z=0$ merging rates in clusters are very low,
but are still relatively high inside groups.}
\label{merging_rate}
\end{figure}

We found that 28\% of the cluster halos, 29\% of the group halos and
52\% of the isolated halos never underwent a major-merging
event. Fig. 1 presents the number of major-merging events -- the
merging rate -- of halos in different environments. We show the
merging rate averaged over three (two at $z>2$) subsequent time
intervals. The error bars are $\sqrt{N}$ errors for the number of
events detected. The three points for merging inside clusters at $z =
0.1,\; 0.25$, and 0.5 come from 1, 3, and 4 events, respectively. The
points in the figure can be fitted by a curve $\alpha \; (1+z)^\beta
\;\exp(\gamma(1+z))$ with $\alpha = -2.1, \; -1.6, \; -1,6 $ and
$\beta = 3.1, \; 2.8, \; 2.5$ for cluster, group, and isolated halos,
respectively. The exponential dilution is the same for all types,
$\gamma = -0.6$. It is mainly due to the fact that at $z > 4$ we are
rapidly loosing the halo progenitors due to mass resolution. The
position and high of the maxima as well as the slope depend slightly
on the chosen threshold for definition of major merging events,
but the relative position of the curves remains practically
constant. 
 
The higher probability of major-merging that cluster and group halos
had in the past is due to the higher density in regions, where cluster
and groups were to form.  Note that clusters have not yet existed at
that time. As clusters with large internal velocities form, merging
rate drastically decreases.  There are almost no major-merging events
of cluster halos in the recent past. Those very few events have
probably happened just outside the clusters before the mergers fell in
the clusters. The probability of recent major-merging is almost the
same for isolated galaxies and for group galaxies. It is very
different in the past: the merging of galaxies, which end in clusters,
was much higher than for isolated galaxies.

\acknowledgements This work was funded by the NSF and NASA grants to
NMSU, and collaborative NATO grant CRG 972148.  SG acknowledges support
from Deutsche Akademie der Naturforscher Leopoldina with means of the
Bundesministerium f\"ur Bildung und Forschung grant LPD 1996.
%% The numerical simulations has been carried out on
%%the Origin2000 computers at NCSA and NRL.

% The endnotes section will be placed here.
\theendnotes

\end{article}
\end{document}